\documentclass[a4paper,11pt]{article}
\pdfoutput=1 

\usepackage{jcappub} 

\usepackage[T1]{fontenc} 
\usepackage{array}

\title{\boldmath A Precise Ultra - Dense Star Model on Spheroidal Space-Time}

\author[a,1]{Hemalatha.R
,\note{Corresponding author.}}
\author[a]{Naren Babu O.V}
\author[b]{Narayanankutty Karuppath}
\author[c]{Sabu M.C}

\affiliation[a]{Amrita Vishwa Vidyapeetham, Department of Mathematics, Amritapuri, 690525,India}
\affiliation[b]{Amrita Vishwa Vidyapeetham, Department of Physics, Amritapuri, 690525,India}
\affiliation[c]{ St. Alberts College (Autonomous),Department of Mathematics, Ernakulam, 682018,India}

\emailAdd{hemalathar@am.amrita.edu}
\emailAdd{narenbabu@am.amrita.edu}
\emailAdd{narayanankuttyk@am.amrita.edu}
\emailAdd{sabuchacko@alberts.edu.in}

\abstract{

This study presents a static, spherically symmetric configuration in which the interior geometry of a relativistic superdense star is modeled as a three-spheroid with constant $t_{1}$. The model is constructed using an analytical closed-form solution to Einstein’s field equations. Assuming a characteristic density of $\rho_{a}= 2\times10^{14} gm.cm^{-3}$, we compute the total mass and radius of the star based on a prescribed set of structural parameters that influence the density profile. The resulting stellar configurations exhibit boundary radii and total masses comparable to those of neutron stars with vanishing charge density.

New exact solutions are obtained by solving the relevant second-order ordinary differential equations. We demonstrate that these solutions satisfy the standard energy conditions and maintain hydrostatic equilibrium throughout the stellar interior. All physical requirements remain valid at every point within the configuration. In this framework, the parameter $\lambda=\frac{\rho_{a}}{\rho_{0}}$ serves as a key determinant of the mass–radius relationship. We further assess the suitability of the model for representing a relativistic superdense star and analyze its stability under radial perturbations. The investigation indicates that, for the configuration to remain dynamically stable, the density ratio between the outer and inner regions must take the value $\lambda$ = 0.4.
}

\begin{document}
\maketitle
\flushbottom

\section{keyword}
Neutron stars , Field equations, Schwarzschild interior solution, Reissner-nordstrom.


\section{Introduction}
\label{sec:introduction}
Einstein’s field equations inherently include the self-interaction of the gravitational field, a feature that greatly restricts the possibility of deriving simple, exact solutions suitable for modelling relativistic stellar interiors. This challenge is compounded by the scarcity of reliable information concerning the behaviour of matter at the extreme densities present in the cores of compact stars, which forces researchers to adopt broad theoretical assumptions regarding their internal physics. For these reasons, analytic models that provide clear, tractable descriptions of relativistic stars are of significant interest \cite{tolman1939static}. Closed-form solutions of this kind \cite{tolman1939static, adler1974fluid, leibovitz1969spherically, PhysRev.116.1027} continue to play a central role in astrophysical studies, especially when they satisfy the essential physical criteria required of ultra-dense matter—such as regularity, physical acceptability, and consistency with relativistic hydrostatic equilibrium.

Vaidya explained that spacetimes with $t_{1}$ constant sections take the form of a 3-spheroid, where $k_{1}$
 quantifies the degree of flatness and $R_{1}$ measures the roundness of the spacetime. These spacetimes are valuable for constructing relativistic models that are easy to analyze, especially for very dense stars like neutron stars. The research by \cite{knutsenh.1988} assessed the physical plausibility of the Vaidya and Tikekar models and demonstrated that these models are stable against small radial pulsations. Another related class of models, exhibiting the same geometry, was introduced by Tikekar \cite{tikekar1990exact}.

Given that only a small number of closed-form analytic solutions to Einstein’s field equations are known for static, spherically symmetric matter distributions—solutions that serve as practical models for superdense stellar objects—it becomes important to explore whether additional classes of models may also prove physically viable. Expanding the range of admissible configurations could offer valuable insights for a variety of intergalactic and stellar systems \cite{tikekarr, jyothy2015diffuse, narayan2017dust, bose2015ultraviolet}.

Ramesh Tikekar and V. O. Thomas studied the geometric properties of the 3D physical spaces derived from $t_{1}$
= constant hypersurfaces of such spacetimes.  In this work, we present a solution to Einstein’s field equations representing the spacetime of a spherically symmetric matter distribution in hydrostatic equilibrium, wherein the physical three-space assumes the geometry of a three-spheroid embedded in a four-dimensional Euclidean space. We analyze the suitability of this solution for modeling the interior structure of highly compact stars, such as neutron stars.

First, consider the curvature parameter $k_{1}$. If $k_{1}$
 is negative, it can represent the energy density of the Vaidya-Tikekar isotropic superdense star, which decreases as one moves away from the core \cite{gupta2000most}. Section 3 discusses the general features of equilibrium matter distributions in a spheroidal space-time. Solutions to Einstein’s field equations are obtained by integrating a second-order ordinary differential equation. The spheroidal parameter $k_{1}$ places a restriction on the existence of new solution classes. The first solution is polynomial in nature. In Section 4, we present a product solution that combines both polynomial and algebraic terms. A specific, stable and precise solution to Einstein’s field equations is reported in this configuration, forming the basis for constructing a relativistic model of a superdense star in later sections.

In Section 5, we apply matching conditions to determine the model parameters after obtaining the solution. Section 6 discusses the stability analysis of the solution and examines how the spheroidal curvature parameter $k_{1}$ influences physical quantities.

In this study, we derive a solution to Einstein’s field equations characterizing the spacetime of a spherically symmetric matter distribution in hydrostatic equilibrium, where the three-dimensional spatial geometry is represented by a three-spheroid embedded in a four-dimensional Euclidean space. We investigate the potential of this solution to model the internal structure of extremely dense stars, such as neutron stars. Section 2 presents the general framework and key properties of equilibrium configurations in spheroidal spacetimes, while Section 3 introduces a robust and analytically precise solution to Einstein’s equations. This solution forms the basis for developing a relativistic model of a superdense star in the subsequent analysis.

Based on \cite{vaidya} and \cite{tikekar1990exact}, the surface density of the star is approximately $2\times10^{14} gm.cm^{-3}$. The corresponding mass and radius are computed for different values of the parameter $\lambda$, which governs the internal density profile. The results confirm that the presented closed-form solution constitutes a physically viable model for relativistic stars, with mass and radius consistent with typical neutron star parameters.
\section{Distribution of Matter in Spheroidal Space-time}

In accordance with \cite{vaidya}, we examine a spherical matter distribution represented as a stationary ideal fluid, with the underlying spacetime defined by the metric

\begin{equation}
\bar{ds}_1^2 = -\left[\frac{1 - k_1 \frac{r_1^2}{R_1^2}}{1 - \frac{r_1^2}{R_1^2}}\right] dr_1^2 - r_1^2 d\theta_1^2 - r_1^2 \sin^2 \theta_1 d\phi_1^2 + e^{\nu(r_1)} dt_1^2,
\label{eqn5}
\end{equation}
where \[k_1 = 1 - \frac{b^2}{R_1^2}.\]
and the 3-space at constant \(t_1\) corresponds to a spheroidal hypersurface embedded in Euclidean 4-space.
The associated 3-space of this spacetime (hypersurface at constant \(t_1\)) has the geometry of a 3-spheroid immersed in a 4-dimensional Euclidean space.

Assuming the matter content is an ideal fluid, the energy–momentum tensor takes the form
For an isotropic fluid, the energy–momentum tensor and four-velocity are given by the standard expressions (see Eqs. (3.5)–(3.6) in \cite{tikekar1990exact}),

\begin{equation}
\bar{T}_{ij} = \left(\rho_1 + \frac{p_1}{c^2}\right) u_i u_j - \frac{p_1}{c^2} g_{ij},
\label{eqn6}
\end{equation}
\begin{equation}
\bar{u}^i = \left(0, 0, 0, e^{-\frac{\nu}{2}}\right).
\label{eqn7}
\end{equation}
where \(p_1\) and \(\rho_1\) denote the fluid pressure and matter density, respectively.
The Einstein field equations yield relations for 
\(\rho(r)\), \(p(r)\) and \(\nu(r)\); the explicit forms of these equations can be found in \cite{vaidya, tikekar1990exact} (their Eqs. (8)–(11)). For completeness, we summarize that the density profile takes the form
\begin{equation}
\frac{8\pi G}{c^2}\rho(r) = \frac{3(1 - k_1)}{R_1^2} \frac{1 - \frac{k_1}{3}\frac{r^2}{R_1^2}}{\left(1 - k_1 \frac{r^2}{R_1^2}\right)^2},
\label{eq:density}
\end{equation}
while the pressure and potential \(\nu(r)\) satisfy the differential constraint derived in \cite{tikekar1990exact} (Eq. (4.1) therein). The hydrostatic equilibrium condition, equivalent to the relativistic Tolman–Oppenheimer–Volkoff equation, is adopted from \cite{tikekar1990exact} (see their Eq. (12)).

This framework eliminates the need for an explicit equation of state; instead, the spheroidal geometry itself dictates the functional behavior of the matter variables.


\section{A Formulation  of the Field Equations}
Equation (7) in \cite{vaidya} constitutes a second-order, nonlinear ordinary differential equation, which is expressed by the introduction of new variables $\psi$ and $z$.

 \begin{equation}
     \psi =e^{\frac{\nu }{2}}   ,   
    \hat z^{2}=\left( 1-\frac{r_{1}^{2}}{R_{1}^{2}} \right),
 \label{eqn14}
 \end{equation}
  
Takes on a basic form 
 \begin{equation}
 {\left(1-k_{1}+k_{1}z^{2} \right)\frac{d^{2}\psi }{dz^{2}}-k_{1}z\frac{d\psi }{dz}  +k_{1}\left( k_{1}-1 \right)\psi}=0
 \label{eqn15}
 \end{equation}
   
 This pertains to a category of linear, second-order ordinary differential equations characterized by the parameter $k_{1}$. Closed-form solutions for specific equations in this category can be derived using appropriate integration techniques. In \cite{patricwils1990}, the equation examined in \cite{tikekar1990exact} has been reformulated as a Riccati equation, and the physical significance of the derived solutions was analyzed for two specific selections of the class parameter.
 
Implementing the transformation for $k_{1}<0$.
 \begin{equation}
     u^{2}=\left( \frac{k_{1}}{k_{1}-1} \right)z^{2}
     \label{eqn16}
\end{equation}

The differential equation (4.2) can be transformed into the form 
\begin{equation}
     {\left(1-u^{2} \right)\frac{d^{2}\psi }{du^{2}}+u\frac{d\psi }{du}+\left( 1-k_{1} \right)\psi} =0,
     \label{eqn17}
 \end{equation}
 This example was examined by \cite{vaidya} for the value $k_{1} = -2$. A comprehensive approach for deriving closed-form solutions to the equation in \cite{tikekar1990exact} is thoroughly examined by \cite{maharajs.dleachp.g.l}. As the complexity of the answers to (4.2) escalates, the surveyability of the models derived from them becomes ever more challenging. Therefore, it is essential to identify manageable models of fluid stars derived from the closed-form solutions presented in \cite{tikekar1990exact}. The standard closed-form solution for the particular selection $k_{1} = -34$ is shown in equation (7) in \cite{vaidya}. By employing the hypergeometric function with \( k_{1}=2-n^{2} \) as a foundation, we may also derive a closed-form solution to this differential equation through an alternative method \cite{chandrasekhar1964dynamical}.

 \begin{eqnarray}
   \bar\psi =As_{1}\left( 1-\frac{204}{35}s_{1}^{2}+ \frac{55488}{6125}s_{1}^{4}-\frac{1257728}{300125}s_{1}^{6}\right)+
     B \left(1-s_{1}^{2}  \right)^{\frac{3}{2}}\left( 1-16s_{1}^{2} +\frac{80}{3}s_{1}^{4}\right)
     \label{eqn18}
 \end{eqnarray}
 here \( s_{1}^2 = \frac{34}{35} z^2 \), and A and B are arbitrary constants of integration. The explicit formulation of the spacetime metric corresponding to this outcome is as follows:

 %
\begin{eqnarray}
\begin{split}
ds_{1}^{2} &= -35\left( \frac{1-s_{1}^{2}}{z^{2}} \right)dr_{1}^{2}-r_{1}^{2}\left( d\theta_{1} ^{2}+sin^{2}\theta_{1} d\phi_{1} ^{2} \right) \\
&+\left[As_{1}\left( 1-\frac{204}{35}s_{1}^{2}+ \frac{55488}{6125}s_{1}^{4}-\frac{1257728}{300125}s_{1}^{6}\right)\right. \\
 & \hspace{4cm} \left.+B \left(1-s_{1}^{2}  \right)^{\frac{3}{2}}\left( 1-16s_{1}^{2} +\frac{80}{3}s_{1}^{4}\right) \right]^{2}dt_{1}^{2}.
 \label{eqn19}
\end{split}
\end{eqnarray}
and the physical variables follow as
\begin{equation}
\frac{8\pi G \rho}{c^2} = \frac{105}{R_1^2}\frac{1 + \frac{34}{3}\frac{r^2}{R_1^2}}{\left[1 + 34\frac{r^2}{R_1^2}\right]^2},
\end{equation}
and
\begin{equation}
\frac{8\pi G p}{c^4} = f(A,B,r_1),
\end{equation}
where $f(A,B,{r_1})$ denotes the algebraic function derived from Eq.~(6) in \cite{vaidya} above.
Further discussion of the physical viability of this model is presented in the subsequent section.

 \section{Physical Interpretation and Validity}
 The derived solution is not contingent upon any assumptions regarding inter-particle interactions. Therefore, it is essential to assess its physical plausibility. Within the scope of validity, the subsequent criteria must be fulfilled:
\begin{enumerate}
    \item Fluid pressure $p_{1}$ and matter density $\rho_{1}$ must be positive at all locations.
    \item The gradients $d\rho_{1}/dr_{1}$ and $dp_{1}/dr_{1}$ must be less than zero.
    \item The speed of sound must always be less than the speed of light, i.e., $dp_{1}/d\rho_{1} < c^{2}$.
    \item The inside metric (Eq.~(4.6)) must seamlessly align with the outer Schwarzschild measure.
\end{enumerate}

The external Schwarzschild line element is expressed as
\begin{equation}
ds_{1}^{2} = -\left(1-\frac{2m}{r_{1}}\right)^{-1}dr_{1}^{2}
- r_{1}^{2}\left(d\theta_{1}^{2} + \sin^{2}\theta_{1} d\phi_{1}^{2}\right)
+ \left(1-\frac{2m}{r_{1}}\right)dt_{1}^{2}.
\label{eq:Schwarzschild}
\end{equation}

From the expression for the density, it follows that $\rho_{1} > 0$ and $d\rho_{1}/dr_{1} < 0$ throughout the configuration. Following the analysis of \cite{tikekar1990exact}, the hydrostatic equilibrium equation (Eq.~(4.2)) implies that, for all $r_{1} < R_{1}$, the pressure gradient satisfies $dp_{1}/dr_{1} < 0$ and the pressure itself remains positive, $p_{1}(r_{1}) > 0$.At the middle ($r_{1} = 0$), the density and pressure reach their respective values.
\begin{equation}
\frac{8\pi G}{c^{2}}\rho_{0} = \frac{105}{R_{1}^{2}},
\qquad
\frac{8\pi G}{c^{4}}p_{0} = \frac{1}{R_{1}^{2}}\left[\frac{-1.30A + 7.87B}{0.04A + 0.05B}\right],
\label{eq:central_density_pressure}
\end{equation}
assuming $A > 0$.

The positivity of pressure at the centre requires
\begin{equation}
-0.8 < \frac{B}{A} < 2.2.
\label{eq:BA_condition1}
\end{equation}
If the strong energy condition $\rho_{0} - 3p_{0}/c^{2} \geq 0$ is further imposed, the range of $\frac{B}{A}$ becomes more restricted:
\begin{equation}
-0.8 < \frac{B}{A} < 0.47.
\label{eq:BA_condition2}
\end{equation}

At the stellar surface ($r_{1}=a$), the pressure is null, $p_{1}(a)=0$. This requirement, along with the continuity of the metric coefficients, produces

\begin{equation}
m = \frac{4\pi G}{c^{2}}\int_{0}^{a}\xi^{2}\rho(\xi)\, d\xi
= \frac{35}{2R_{1}^{2}}\frac{a^{3}}{\left[1+34(a^{2}/R_{1}^{2})\right]}.
\label{eq:mass}
\end{equation}

The corresponding matching condition between the interior and exterior metrics is expressed as
$$\begin{array}{ll}
    \left( 1-\frac{2m}{a} \right)^{1/2} = & \\ 
    \frac{-2As_{1}\left (\frac{37}{2}-\frac{4182}{35}s_{1}^{2}+\frac{249696}{1225}s_{1}^{4}-\frac{628864}{6125}s_{1}^6\right )-2B\left ( 1-s_{1}^{2} \right )^{\frac{1}{2}}\left(\frac{35}{2}-\frac{659}{2}s_{1}^{2}+\frac{2656}{3}s_{1}^{4}-\frac{1720}{3}s_{1}^{6}\right )}{35R_{1}^{2}\left (1-s_{1}^{2} \right )\left [As_{1}\left( 1-\frac{204}{35}s_{1}^{2}+ \frac{55488}{6125}s_{1}^{4}-\frac{1257728}{300125}s_{1}^{6}\right)+B \left(1-s_{1}^{2}  \right)^{\frac{3}{2}}\left( 1-16s_{1}^{2} +\frac{80}{3}s_{1}^{4}\right))\right ]}
    \label{eqn28}
\end{array} $$
\begin{equation}
    ~
\end{equation}

The condition $p_{1}(a) = 0$ leads to

\begin{eqnarray}
\nonumber -2As_{1}\left (\frac{37}{2}-\frac{4182}{35}s_{1}^{2}+\frac{249696}{1225}s_{1}^{4}-\frac{628864}{6125}s_{1}^6\right )=\\
2B\left ( 1-s_{1}^{2} \right )^{\frac{1}{2}}
 \left(\frac{35}{2}-\frac{659}{2}s_{1}^{2}+\frac{2656}{3}s_{1}^{4}-\frac{1720}{3}s_{1}^{6}\right )
 \label{eqn29}
\end{eqnarray}

which, together with Eq.~(5.6) and (5.7), determines the constants $A$ and $B$, whereas Eq.~(\ref{eq:mass}) provides the total mass $m$ of the system.
The parameter for density variation is specified as
\begin{equation}
\lambda = \frac{\rho_{a}}{\rho_{0}}
= \frac{8\pi G\rho_{1}}{c^{2}}
= \frac{\left[1+\frac{34}{3}(a^{2}/R_{1}^{2})\right]}
{\left[1+34(a^{2}/R_{1}^{2})\right]^{2}},
\label{eq:lambda}
\end{equation}
where $\rho_{a}$ and $\rho_{0}$ denote the densities at the boundary and at the centre, respectively. For given values of $\rho_{a}$ and $\lambda$, Eqs.~(5.2) and (5.8) determine $R_{1}^{2}$ and $a^{2}/R_{1}^{2}$, and thus $a$ and $R_{1}$. Equation (5.6) then fixes the total mass $m$.

Following \cite{tikekar1990exact,naren2020exact,sabu2021static,sasidharan2020general}, the expression for the sound speed (for $k_{1}=-34$) is obtained as
\begin{eqnarray}
\frac{dp_{1}}{d\rho_{1}} =
\frac{
\pi G R_{1}^{2}\left(\rho_{1}+\frac{p_{1}}{c^{2}}\right)
\left[35+\left(8\pi \frac{G}{c^{4}}\right)p_{1}R_{1}^{4}\left(1+34\frac{r_{1}^{2}}{R_{1}^{2}}\right)\right]
\left(1+34\frac{r_{1}^{2}}{R_{1}^{2}}\right)^{3}
}{
595\left(1-\frac{r_{1}^{2}}{R_{1}^{2}}\right)\left(5+34\frac{r_{1}^{2}}{R_{1}^{2}}\right)}.
\label{eq:dpdrho}
\end{eqnarray}

At the centre, the strong energy condition $\rho_{0}-3p_{0}/c^{2}\geq 0$ implies
\begin{equation}
\left[\frac{dp_{1}}{d\rho_{1}}\right]_{r_{1}=0} < 0.41\,c^{2}.
\label{eq:dpdrho_centre}
\end{equation}
At the boundary ($r_{1}=a$),
\begin{equation}
\left[\frac{dp_{1}}{d\rho_{1}}\right]_{r_{1}=a}
= \frac{105\left(1+34a^{2}/R_{1}^{2}\right)
\left(1+34a^{2}/3R_{1}^{2}\right)}
{136\left(1-a^{2}/R_{1}^{2}\right)
\left(5+34a^{2}/R_{1}^{2}\right)}\,c^{2},
\label{eq:dpdrho_boundary}
\end{equation}
which remains less than $c^{2}$ if $a/R_{1} < 0.46$. This requirement is satisfied for all models in Table~1 that obey the strong energy condition. Since $\frac{dp_{1}}{d\rho_{1}}$ does not vanish anywhere within $r_{1}\leq a$, it follows that $\frac{dp_{1}}{d\rho_{1}} < c^{2}$ throughout the configuration. Consequently, the sound speed remains subluminal everywhere inside the stellar model.

The values of the relevant physical parameters $\lambda$, $R_{1}$, $r_{1}$, $M$, $M/M_{\odot}$, $A$, and $B$ corresponding to the models studied are summarized in Table~1.

             \label{eqn27}
    ~

\begin{table*}[!ht]
\centering
\vspace{0.7cm}

\begin{tabular}{cccccccccc}
\hline
S.No & $\lambda$ & $R_{1}(km)$ & $a(km)$ & $a/R _{1}$ & $\frac{M}{M_{\bigodot }^{a}}$ & $A$ & $B$ & $M(km)$ & $B/A$    
\\
\hline 
1  & 0.95 & 163.68 & 4.97  & 0.03 & 0.05 & 20.76  & -0.03 & 0.07 & -0.001 \\
2  & 0.9  & 159.31 & 6.99  & 0.04 & 0.15 & 19.32  & -0.04 & 0.22 & -0.001 \\
3  & 0.85 & 154.82 & 8.53  & 0.06 & 0.28 & 18.09  & -0.04 & 0.41 & -0.002 \\
4  & 0.8  & 150.20 & 9.79  & 0.07 & 0.43 & 17.03  & -0.04 & 0.63 & -0.002 \\
5  & 0.75 & 145.43 & 10.89 & 0.07 & 0.61 & 16.14  & -0.04 & 0.81 & -0.002 \\
6  & 0.7  & 140.50 & 11.87 & 0.08 & 0.81 & 15.43  & -0.04 & 1.19 & -0.003 \\
7  & 0.65 & 135.39 & 12.74 & 0.09 & 1.03 & 14.89  & -0.04 & 1.52 & -0.003 \\
8  & 0.6  & 130.17 & 13.53 & 0.10 & 1.27 & 14.59  & -0.05 & 1.87 & -0.003 \\
9  & 0.55 & 124.54 & 14.25 & 0.11 & 1.53 & 14.61  & -0.05 & 2.26 & -0.004 \\
10 & 0.5  & 118.74 & 14.90 & 0.13 & 1.81 & 15.13  & -0.06 & 2.67 & -0.004 \\
11 & 0.45 & 112.65 & 15.49 & 0.14 & 2.11 & 16.66  & -0.07 & 3.11 & -0.004 \\
12 & 0.4  & 106.21 & 16.02 & 0.15 & 2.43 & 20.83  & -0.09 & 3.58 & -0.004 \\
13 & 0.35 & 99.35  & 16.49 & 0.17 & 2.78 & 38.39  & -0.17 & 4.10 & -0.004 \\
14 & 0.3  & 91.98  & 16.89 & 0.18 & 3.15 & -77.26 & 0.31  & 4.65 & -0.004 \\
15 & 0.25 & 83.96  & 17.22 & 0.21 & 3.54 & -12.19 & 0.03  & 5.22 & -0.003 \\
16 & 0.2  & 75.10  & 17.47 & 0.23 & 3.95 & -4.86  & -0.01 & 5.83 & 0.002  \\
17 & 0.15 & 65.04  & 17.62 & 0.27 & 4.34 & -2.34  & -0.04 & 6.40 & 0.028  \\
18 & 0.10 & 53.10  & 17.61 & 0.33 & 4.85 & -0.85  & -0.18 & 7.15 & 0.217\\
 \hline
\end{tabular}
\caption{
Body of materials and balance radii corresponding to $\rho_{a}=2\times10^{14} \text{ gm.cm}^{-3}$, for a group of relativistic star models.
Here  $M=\frac{mc^{2}}{G}, M_{\bigodot }^{a}$= mass of the sun.
}
\label{fig1}
\end{table*}

\section{Stability Properties of the Configuration}

Chandrasekhar \cite{chandrasekhar1964dynamical} developed a perturbative framework to examine the dynamical stability of relativistic star configurations in response to infinitesimal radial perturbations. This method delineates a standard mode of radial oscillation for a static equilibrium arrangement.

\begin{equation}
    \delta r_{1} = \xi(r_{1})_{\text{trial}} \, e^{i \sigma t_{1}},
    \label{eqn34}
\end{equation}
which corresponds to a stable oscillation if and only if the eigenfrequency \(\sigma\) is real \cite{knutsenh.1988}.

In accordance with \cite{bardeen1966catalogue}, we define the trial function \(u\) as

\begin{equation}
    u = \xi_{\text{trial}} \, r_{1}^{2} \, e^{-\nu/2},
    \label{eqn35}
\end{equation}
where \(\nu = \nu(r_{1})\) is the gravitational potential function in the static metric.

For a spherically symmetric line element of the form
\begin{equation}
    ds_{1}^{2} = e^{\nu(r_{1})} dt_{1}^{2} - e^{\lambda(r_{1})} dr_{1}^{2} - r_{1}^{2} \left( d\theta_{1}^{2} + \sin^{2} \theta_{1} \, d\phi_{1}^{2} \right),
    \label{eqn36}
\end{equation}
Chandrasekhar's pulsation equation reads
\begin{align}
\sigma^2 \int_{0}^{a} e^{(3\lambda + \nu)/2} \left(p_{1} + \rho_{1}\right) \frac{u^2}{r_{1}} \, dr_{1}
= \int_{0}^{a} e^{(3\lambda + \nu)/2} \frac{\left(p_{1} + \rho_{1}\right)}{r_{1}^{2}} \Bigg\{
&\left[ -\frac{2}{r_{1}} \frac{d\nu}{dr_{1}} - \frac{1}{4} \left( \frac{d\nu}{dr_{1}} \right)^2 + 8\pi p_{1} e^{\lambda} \right] u^2 \nonumber \\
&+ \frac{dp_{1}}{d\rho_{1}} \left( \frac{du}{dr_{1}} \right)^2 \Bigg\} dr_{1},
\label{eqn37}
\end{align}
where \(a\) is the boundary of the fluid distribution and \(\frac{dp_{1}}{d\rho_{1}}\) represents the adiabatic speed of sound squared.

We impute the condition that the Lagrangian perturbation of the pressure is zero at the surface (\(r_{1} = a\) yields
\begin{equation}
    \Delta p_{1} = - e^{\nu/2} \frac{\gamma p_{1}}{r_{1}^{2}} \frac{du}{dr_{1}} = 0,
    \label{eqn38}
\end{equation}
which implies
\begin{equation}
    \left. \frac{du}{dr_{1}} \right|_{r_{1}=a} = 0.
    \label{eqn39}
\end{equation}

Utilising the methodology delineated in \cite{bardeen1966catalogue} and subsequently embraced by \cite{knutsenh.1988}, we execute the integration of the pulsation equation employing a trial function of the specified form

\begin{equation}
    u = R_{1}^{3} \left(1 - z^{2} \right)^{3/2} \left[ 1 + a_{1}(1 - z^{2}) + b_{1}(1 - z^{2})^{2} + \cdots \right],
    \label{eqn40}
\end{equation}
where \(a_{1}, b_{1}, \ldots\) are arbitrary constants, and \(z = r_{1}/R_{1}\) is a dimensionless radial variable. The boundary condition in this form becomes
\begin{equation}
    3 + 5 a_{1} y^{2} + 7 b_{1} y^{2} + \cdots = 0,
    \label{eqn41}
\end{equation}
where \(y = 1 - z^{2}\).

The integral contributing to the right-hand side of the pulsation equation, labelled \(I_{R_{1}}\), is given by
\begin{align}
    I_{R_{1}} = \frac{R_{1}}{8\pi \sqrt{2}} \int_{0}^{a} e^{(\lambda + 3\nu)/2} \frac{(p_{1} + \rho_{1})}{R_{1}^{2}} y
    \Bigg\{ &y \left[ -\frac{2}{z} \frac{d\nu}{dz} - \frac{y}{4z^{2}} \left( \frac{d\nu}{dz} \right)^{2} + 8\pi p_{1} e^{\lambda} \right] \nonumber \\
    &\times \left( 1 + a_{1} y + b_{1} y^{2} \right)^{2}
    + \frac{dp_{1}}{d\rho_{1}} \left( 3 + 5a_{1} y + 7b_{1} y^{2} \right)^{2} \Bigg\} \sqrt{\frac{1}{y}} \, dy,
    \label{eqn42}
\end{align}
where the metric functions \(\nu\), \(\lambda\), and the fluid variables \(p_{1}, \rho_{1}\) are taken from equations Eqs.~(4.5), (4.7) and (4.8) in  \cite{tikekar1990exact} and evaluated accordingly.

By selecting suitable combinations of the parameters \(a_{1}\) and \(b_{1}\), we ascertain that \(I_{R_{1}} > 0\) for many physically feasible configurations outlined in Table 1. This offers compelling evidence that the models examined are dynamically stable about miniscule radial pulsations. Table 2 displays the values of \(I_{R_{1}}\) for the selection \(b_{1} = 1\) and \(a_{1} = \frac{3 + 7 y^{2}}{5y}\).



\begin{table}[!ht]
\vspace{0.5cm}
\centering
\begin{tabular}{ccll}
\hline
S.No & $a_{1}$ & $a(km)$ & \hspace{2cm}$I_{R_{1}}$\\
\hline
1 & 3.79 & 0.169 & 1.22332773526112$\star10^{17}$ \\
2 & 2.25 & 0.338 &  3.1682369891969$\star10^{18}$\\
3 & 1.89 & 0.507 &  6.76592647111822$\star10^{18}$ \\
4 & 1.83 & 0.676 & 1.27212885904473$\star10^{19}$  \\
5 & 1.89 & 0.845 &  2.38161320437971$\star10^{19}$ \\
6 & 2.01 & 1.014 &  4.5228548399543$\star10^{19} $ \\
7 & 2.16 & 1.183 &  8.63356477061169$\star10^{19}$ \\
8 & 2.34 & 1.352 &  1.63115811781686$\star10^{20}$ \\
9 & 2.52 & 1.521 &  2.88670823695058$\star10^{20}$ \\
10& 2.73 &	1.699& 4.65789691985662$\star10^{20}$ \\
\hline
\end{tabular}
\caption{For the parameter set  $a_{1}, b_{1} = 1$ together with $\lambda = 0.4$, the integrals that appear on the right-hand side of the pulsation equation (\ref{eqn42}) are evaluated to obtain the corresponding pulsation characteristics.}
\end{table}

 \begin{figure}[!ht]
   \centering
   \includegraphics[width=0.4\textwidth]{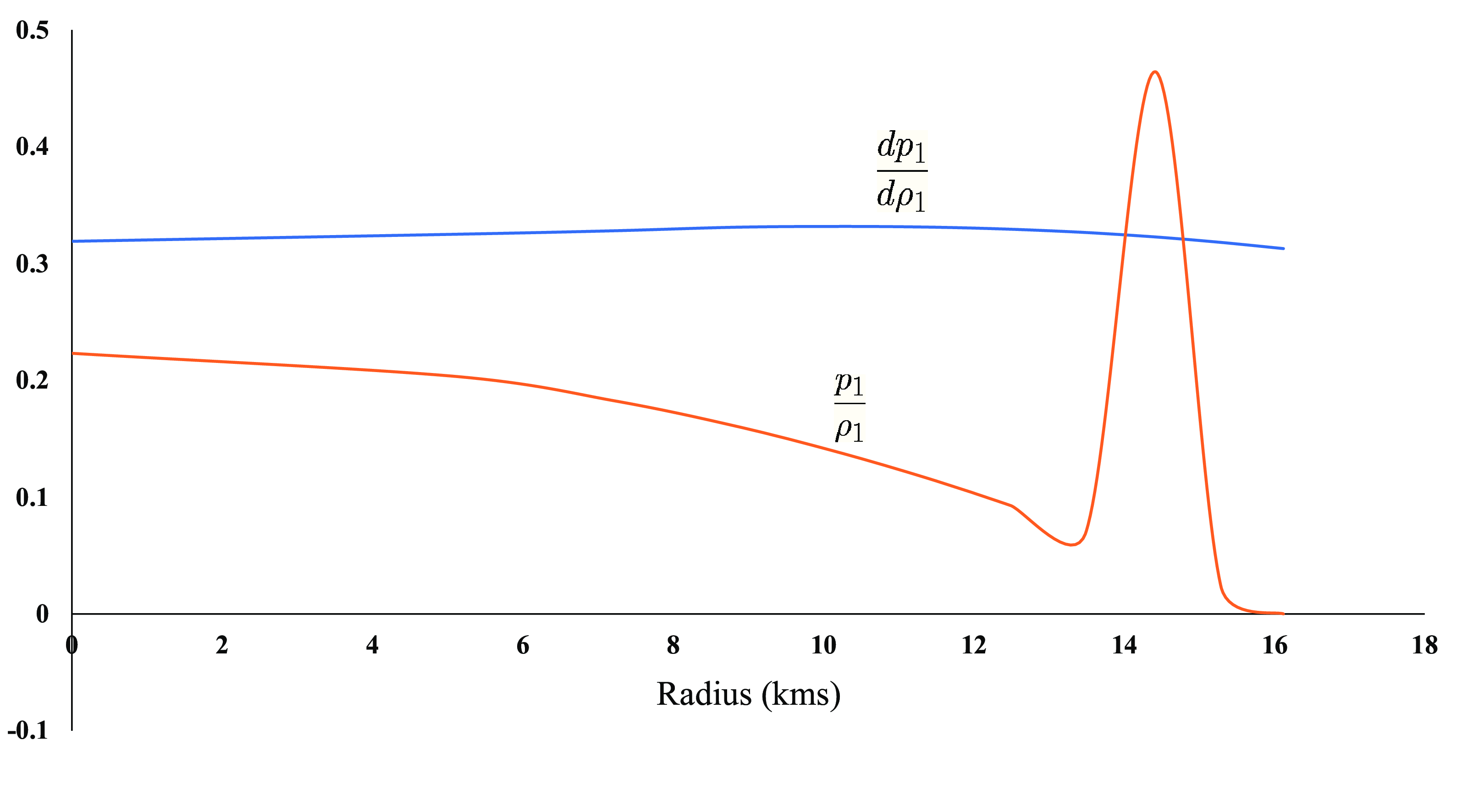}
   \hfill
   \includegraphics[width= 0.4\textwidth]{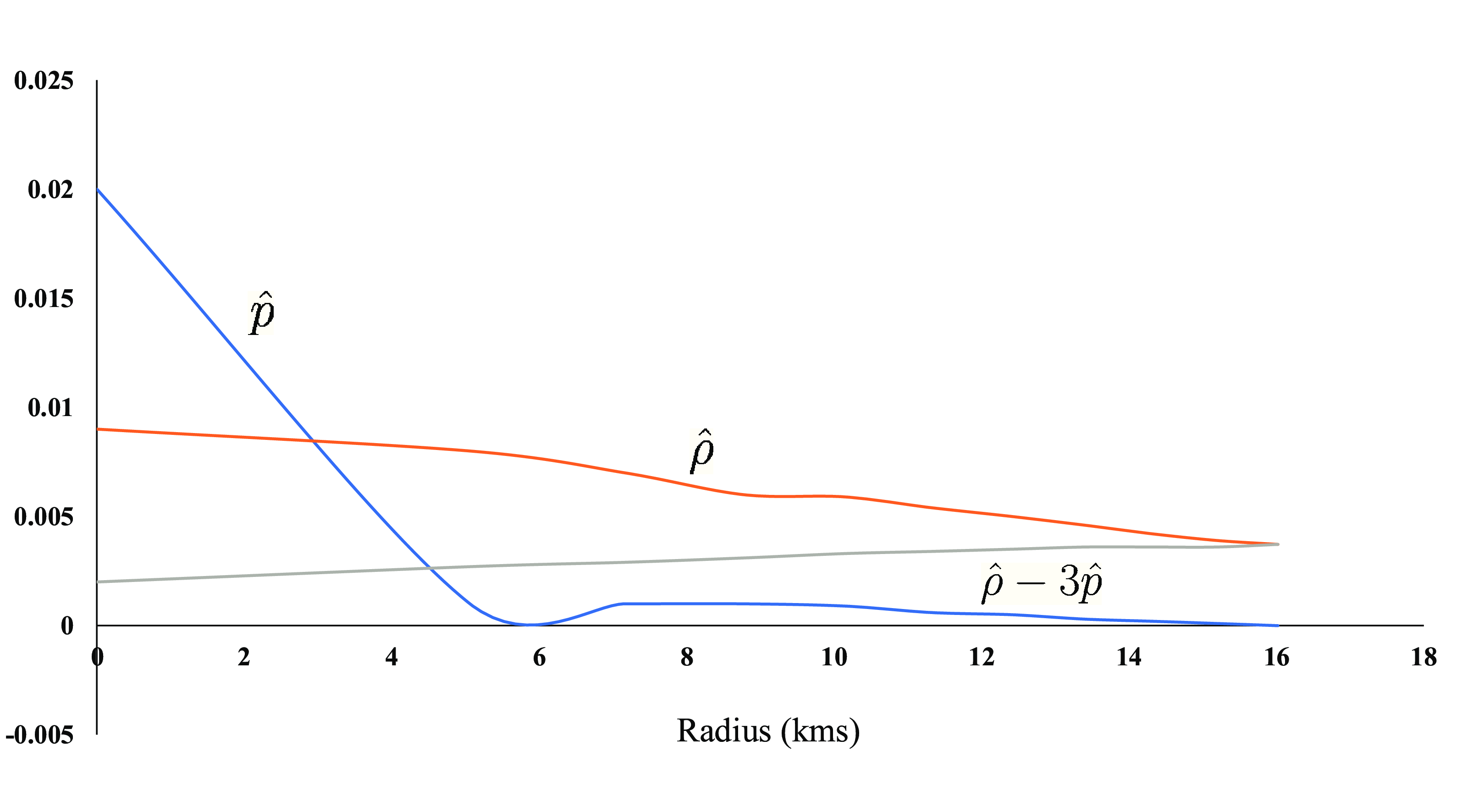}
    
   \caption{(a) Graph of $\hat{\rho }$=$\frac{8G\pi}{c^{2}}\rho_{1}\times10^{4}$  against radius $\hat{p}=\frac{8G\pi}{c^{4}}p _{1}\times10^{4}$ }
   {(b) Graph of $\hat{\rho }$=$\frac{8G\pi}{c^{2}}\rho _{1}\times10^{4}$  and  $\hat{p}$=$\frac{8G\pi}{c^{4}}p _{1}\times10^{4}$against radius $\hat{\rho }-3\hat{p}$}
   
   \label{fig:Fig}
\end{figure}

\begin{figure}[!ht]
    \centering
    \includegraphics[width= 2in]{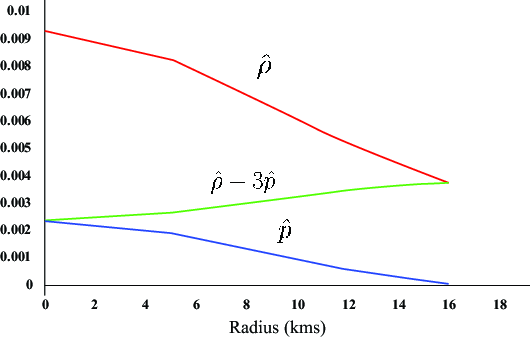}
  \caption{$\lambda$ =0.4 , $k$=-34 The functions $\hat{\rho}$, $\hat{p}$, and $\hat{\rho} - 3\hat{p}$ are plotted as functions of the radial coordinate $r$.}

    \label{fig:Fig.eps3.eps}
\end{figure}

\section{Discussion}
When a star exhausts its thermonuclear energy sources, it begins to contract under its own gravity, eventually transforming into a black hole, neutron star, or white dwarf, depending on its total mass. The models proposed by Vaidya and Ticker [5] describe the formation of a superdense star in the final stages of stellar evolution. Such stars exhibit matter densities Between $10^{14}$ and $10^{16},\mathrm{g.cm^{-3}} $.

The matter density is fixed at $\rho_{a}=2\times10^{14}\mathrm{gm,cm^{-3}}$ near the stellar border $r_{1}=a$, a value typical of neutron-star interiors (see Figure 1). We find the equivalent values of $R_{1}$, $a$, $m$, $B$, and $A$ by applying the numerical method described in Section 3 and methodically changing the parameter $\lambda$. Table 1 compiles these findings.

Our study shows a clear trend: the mass and radius of the star configuration decrease as the density-gradient parameter $\lambda$ grows. Greater values of $\lambda$ result in a stronger central concentration of matter, which is reflected in this behaviour. The anticipated masses and radii also fall within the usual range of observed neutron stars for all acceptable equilibrium models, proving the physical feasibility of the solutions.

Numerical study further substantiates that all models with $\lambda \geq 0.1$ fulfil the physical criteria $\rho_{1}>0$, $p_{1}\geq 0$, and $\rho_{1}-\frac{3p_{1}}{c^{2}}\geq 0$ across the whole distribution.

Astrophysical investigations indicate that the equation of state for neutron matter is adequately understood up to a reference density of $\rho_{f}=5.1\times10^{14},\mathrm{g,cm^{-3}}$. Therefore, the model with $\lambda=0.4$ is notably significant. Figure 1(a) exhibits the variation of $\frac{8\pi G\rho_{1}}{c^{2}}$, $\frac{8\pi Gp_{1}}{c^{4}}$, and $\frac{3\times8\pi Gp_{1}}{c^{4}}$ as a function of the radial coordinate for this scenario, whereas Figure 1(b) demonstrates the associated pressure-density relationship. The configuration reaches its maximum mass at a radius of $1.699,\mathrm{km}$ for $\lambda=0.4$. Relaxing the criterion $\rho - 3p/c^{2} \geq 0$ to $\rho - 3p/c^{2} > 0$ permits models with greater radii and increased masses, contingent upon the fulfilment of condition (6.4).

Consequently, the space–time metric presented in Eq. (5.6) produces a series of equilibrium configurations with surface densities, masses, and radii analogous to those of conventional neutron stars. The current class of stellar models demonstrates reduced mass and radius compared to those associated with $k_{1}=-2$, $k_{1}=-7$, $k_{1}=-14$, and $k_{1}=-23$ with an equivalent density-variation parameter $\lambda$. Models with $k_{1}=-34$ permit increased density variation. To ensure stability, the outer-to-inner density ratio must be 0.4 for $k_{1}=-34$, as illustrated in Figure 2.

\section*{Acknowledgment} 
\addcontentsline{toc}{section}{Acknowledgment}
The writers convey their appreciation to H.H. Mata Amritanandamayi Devi (Amma), Chancellor, for the assistance and resources provided. The authors extend their appreciation to the referees for their vital criticism and to the Director of IUCAA, Pune, for the facilities provided in concluding this work.



\begin{thebibliography}{99}

\bibitem{tolman1939static}Tolman, Richard C, 
  \textit{Static solutions of Einstein's field equations for spheres of fluid}, Physical Review, 55, 4, 364, 1939. 
  
\bibitem{adler1974fluid}Adler, Ronald J
  \textit{A fluid sphere in general relativity}, Journal of Mathematical Physics, 15, 6,727--729, 1974.
   
\bibitem{leibovitz1969spherically}Leibovitz, Clement,
 \textit{Spherically Symmetric Static Solutions of Einstein's Equations},Physical Review, 185, 5, 1664, 1969.
 
\bibitem{PhysRev.116.1027}Buchdahl, H. A.
 \textit{General Relativistic Fluid Spheres}, Phys. Rev., 116, 4, 1027--1034, 0, 1959.
 
\bibitem {knutsenh.1988}Knutsen, H.,
   \textit Mon. Not. R. Astron. Soc., 232, 163, 1988.
   
\bibitem{tikekar1990exact}Tikekar, Ramesh
\textit{Exact model for a relativistic star},
  Journal of mathematical physics, 31, 10, 2454--2458, 1990.

\bibitem {tikekarr}Tikekar,R., and Sabu,M.C.
    \textit New direction in Relativity\& cosmology, 199, 207, 1997.

  
  \bibitem{jyothy2015diffuse}Jyothy, SN and Murthy, Jayant and Karuppath, Narayanankutty and Sujatha, NV
  \textit{Diffuse Radiation from the Aquila Rift},Monthly Notices of the Royal Astronomical Society, 454,2, 1778--1784,2015.

  \bibitem{narayan2017dust}Narayan, Sathya and Murthy, Jayant and Karuppath, Narayanankutty
  \textit{Dust scattering from the Taurus Molecular Cloud},Monthly Notices of the Royal Astronomical Society, 466, 3, 3199--3205, 2017.

  \bibitem{bose2015ultraviolet}Bose, Lakshmi S and Sujatha, NV and Narayanankutty, K
  \textit{Ultraviolet and infrared correlation studies in Orion},Open Astronomy, 24, 3, 319--326, 2015.

  \bibitem{gupta2000most}Gupta, YK and Jasim, MK
  \textit{On most general exact solution for Vaidya-Tikekar isentropicsuperdense star},Astrophysics and Space Science , 272, 403--415, 2000.

  
\bibitem{vaidya}Vaidya, PC and Tikekar, Ramesh
  \textit{Exact relativistic model for a superdense star},
Journal of Astrophysics and Astronomy, 3,3, 325--334, 1982.
   
\bibitem {patricwils1990} Patric Wils
    \textit GRG, 5, 539, 1990.
    
\bibitem{maharajs.dleachp.g.l}Maharaj, S.D and Leach, P. G. L
    \textit J. Math. Phys.(pre print),

    \bibitem{chandrasekhar1964dynamical}Chandrasekhar, S
  \textit{Dynamical instability of gaseous masses approaching the Schwarzschild limit in general relativity}, Physical Review Letters , 12, 4, 114, 1964.

\bibitem{naren2020exact}Naren Babu, OV and Hemalatha, R and Sabu, MC
  \textit{An exact super dense star model on spheroidal space-time},Astrophysics and Space Science, 365, 5, 82, 2020.

  \bibitem{sabu2021static}Sabu, MC and Sasidharan, Aiswarya S and Mattam, Jeet Kurian and others
  \textit{Static General Solution To Einstein's Field Equations},JournalNX , 7, 06, 353--357, 2021.
  
\bibitem{sasidharan2020general}Sasidharan, Aiswarya S and Sabu, MC
  \textit{General solution to Vaidya-Tikekar metric},International Journal of Mathematics Trends and Technology, 67, 08, 15--25, 2021.
   
\bibitem{bardeen1966catalogue}Bardeen, James M and Thorne, Kip S and Meltzer, David W
  \textit{A catalogue of methods for studying the normal modes of radial pulsation of general-relativistic stellar models}, 145, 505, 1966.
  
   
  


\end{thebibliography}
\end{document}